# Market-Based Coordination of Price-Responsive Demand Using Dantzig-Wolfe Decomposition Method

Foad Najafi[1], Matthias Fripp

Department of Electrical Engineering, University of Hawaii at Manoa, 2540 Dole Street, Honolulu, Hawaii

*Abstract*—With the increased share of Distributed Generation (DG) and Demand Responsive (DR) loads in the power systems, new approaches based on the game theory framework have been proposed to tackle the problem of coordination of Price Responsive Devices (PRD). The PRDs are modeled as self-benefiting players who try to optimize their consumption based on the price. In this paper, for the first time, a new algorithm based on the Dantzig-Wolfe (DW) Decomposition method to solve the coordination problem of self-benefiting PRDs in a distributed fashion has been proposed. By utilizing the distributed nature of Dantzig-Wolfe, the PRD's self-benefiting algorithms are modeled as sub-problems of the DW, and the coordinator (or the grid operator) who collects energy consumption of PRDs (their energy bid), solves the master problem of the DW and calculate the price signal accordingly. The proposed algorithm is fast since the subproblem in DW (which could be millions of PRDs) can be solved simultaneously. Furthermore, based on the DW theory, if the PRDs subproblems are convex, reaching the optimal point (Equal to Nash Equilibrium) in limited iterations is guaranteed. A simulation with 8 participant households has been conducted to evaluate the model. Each house is equipped with two types of loads: an Electric Vehicle (EV) as a sample of interruptible loads and an Electric Water Heater (EWH) as a sample of Thermostatically Control Loads (TCA). The results show that when the algorithm reaches the optimal point, the generation cost and the user payment (based on the marginal cost of generation) decrease. Furthermore, the aggregate load's Peak to Average (PAR) reduces significantly.

## 1. Introduction

New renewable and non-renewable Distributed Generation (DG) sources are being added to the power systems daily. Utility-scale storage systems and Demand Responsive (DR) loads are also added as well. Alongside the addition of these modern generation/consumption units, the emergence of fast and reliable communication technologies such as 5G provided a two-way communication path between the generation and consumption sides in power systems. These changes created unprecedented challenges and opportunities in power systems[1], [2]. To overcome these challenges and utilize the new opportunities, the joint management of the demand alongside the generation has been proposed. The co-management of demand/generation has numerous advantages for modern smartgrids, such as balancing supply and demand, integrating more significant portions of renewable energy, and adding auxiliary services such as frequency and voltage provisioning by using the DGs and DRs.

Numerous strategies have been proposed for the joint coordination of appliances in the literature[3], [4]. Among these strategies, the collective coordination of price-responsive devices (PRD) is gaining momentum. The PRD coordination is a market-based approach. On top of having the advantages mentioned above, a market-based approach can incentivize both the generation and consumption side to help achieve the smartgrid paradigm goals. In other words, PRD coordination can bring all the advantages of free market systems defined in economics to energy management systems. Some of these advantages are incorporating a larger share of renewables by creating a market for the excess energy, stabilizing the grid by providing ancillary services such as voltage and frequency provisioning peak shedding, and reducing generation cost, etc. [5], [6]. One of the strategies to coordinate PRDs is through central controllers[7], [8]. In these approaches, the controller sends the control signal to the appliances to achieve a specific objective, such as valley filling or generation cost minimization.

One primary division in PRD coordination is centralized vs. distributed coordination[9]. The central control approaches were beneficial because there was no effective communication method nor price or demand-responsive appliances in the grid in the past. However, with the increased share of PRDs such as batteries, EVs, and TCAs, the central control approach will lose the chance of finding the optimal, and the chance to create a free market where supply and demand negotiate freely will be lost. To overcome these undesirable outcomes, many studies proposed distributed coordination of flexible demands [10], [11].

---

[1] Corresponding Author Email: fnajafi@hawaii.edu

## 1.1 The paradigm of this work

in this paper, we develop a distributed market-based appliance coordination algorithm. The PRDs are modeled as privately rational individuals who try to minimize local costs (solve their private optimization problem). The PRDs consist of Electric Vehicles (EVs) as a sample of interruptible loads and Electric Water Heaters (EWH) as a sample of Thermostatically Controlled Appliances (TCA). The appliances only communicate with an aggregator by sending their consumption plan (energy bid) and receiving the corresponding price signal. The appliances do not share any information with each other, and we assume they do not have market power individually or create a coalition to manipulate the algorithm in their favor.

The main novelties of this work are:

(a) factor in private preferences for good service vs. cost reductions, (b) coordinate devices like EVs and customer-sited batteries that may have an all-or-nothing response to prices (and are likely to be on the margin often, so they are essential to coordinate), and (c) can scale to incorporate any number of devices with any valid price response.

## 2. Related Works

This section reviews the methods that tackle appliance coordination problems where each individual response is essential in decision-making. These methods could have different objectives, such as generation cost minimization, valley filling, or peak shaving. While these objectives are different, the end results are closely correlated. I.e., pursuing an optimal answer for one objective will yield a near-optimal solution for the other objective. E.g., seeking the generation cost minimization [12] yields a flat load curve (valley filling or demand reduction) [13]–[17]. While grid operators can have different objectives which could be closely correlated, as mentioned in [18], [19], the demand-responsive loads can be modeled in two different ways: 1- following the operator objective [20] (direct load control (DLC)). 2- following self-benefiting optimization algorithms [12] in a market-based fashion to bid for energy based on the price signal.

While pursuing the grid operator objective can have direct benefits for the operator (reduced generation costs), it can also have *indirect* benefits for the end-user as a form of discount or credit. However, while these indirect benefits could be beneficial, they may not be optimal for the end user. This need for more optimality to gain the maximum benefit can prevent end-users from joining such programs.

Therefore, the market-based approaches appeal more to the end-user since they give them the ultimate benefit and freedom of choice. One of the leading chains of literature when it comes to distributed market-based coordination of PRDs while they are modeled as self-benefiting and competing units is based on the game theory framework[12], [17], [21]–[24].

[25], [26] are based on Mean Field Game (MFG) approach. In the MFG methods, the individual response of the agents is optional in decision-making. However, the critical factor for decision-making is based on all agents' overall (mean) response. Each appliance responds to the price signal in this method, and the appliance population is approximated as infinite. They showed that their model could reach ε-Nash equilibrium when a quadratic term is used in the cost function of appliances.

In [27], the authors propose a distributed method using Mean Field Game (MFG) to schedule a large population of thermostatically controlled loads (TCLs). The price-responsive rational agents plan their energy consumption and participate in the frequency provisioning market.

Gan et al. [17] proposed a decentralized optimization model for valley-filling using the elastic energy need of EVs. They formulated a scheduling algorithm for EVs whose objective is to fill the valley. The EVs respond to the control signal algorithm. The algorithm is solved iteratively until an optimal solution is

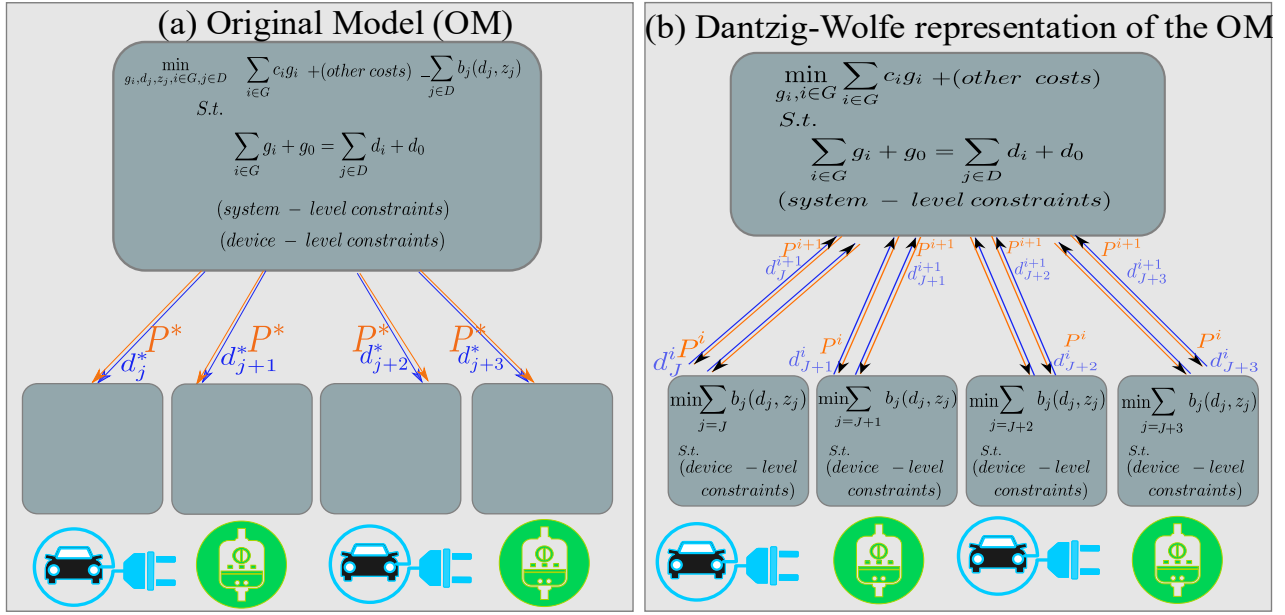

*Figure 1(a) original unit commitment with PRD (solved in one step) (b) Dantzig-Wolfe decomposition of the original model (two iterations of the algorithm)*

obtained. The issue with this method is that it needs to be market-based. i.e., the appliances (here EVs) are supposed to follow the grid objective (valley-filling), not their benefit, i.e., minimizing their own costs. Instead, the individual users follow the operator's objective, which is valley filling. While this could be advantageous for the users overall, it was not shown how each user would benefit from using this method, nor was it shown if the energy cost would decrease by using this approach.

Ref. [23] proposed a distributed method to reach Nash equilibrium (NE) between self-serving plug-in electric vehicles (PEVs). To reach NE, the authors proposed an iterative algorithm where the PEVs respond to a price signal generated based on the average bid of all other PEVs in each iteration. The goal is to make the load pattern of each EV close to all the other EVs. However, while it is assumed that EVs are cost-minimizing, it needs to be clarified how they minimize the generation cost or user payments.

Ref. [28] proposes a decentralized algorithm for PEV control that finds a balance between generation cost and the cost associated with overloading and battery degradation. It was shown that under mild conditions, the algorithm converges to a socially accepted coordination between PEVs. An instantaneous billing scheme to shift the peak-time load is proposed in [29]. The appliances are modeled as selfish units. The authors proposed different methods for different scenarios to reach NE. e.g., when loads are controlled, centralized, or distributed.

Rivera et al. [19] proposed a distributed optimization method for scheduling EVs based on the Alternating Direction of Multipliers (ADMM). They offered a generic multi-objective optimization platform where the cost function is a weighted sum of aggregator and local user objectives. Using this generic platform, they solve the EV coordination problems with two different objectives. i.e., valley-filling and cost minimization. The size of the problem expands linearly with the increment in the size of the EV's fleet. However, the proposed multi-objective is a weighted sum of the end-users and operator's goals. Therefore, based on the value of weights, an unfair bias could be enforced upon either the end-users or the operator.

Using the ADMM method proposed in [19], authors in [30] proposed a method for coordination of EV fleet where local objective optimally schedules EV's charging session and on a higher level, the Macro Load Area Aggregator (MLAA) provides the DER with generation profiles. The method reaches Walrasian equilibrium [31]at its optimum.

Depaola et. Al [12] proposed an iterative PRD coordination based on a game theory framework. They proved that their method reaches the optimal point (Nash Equilibrium) in theory. They demonstrated that at each iteration, the generation cost of electricity decreases. However, their method is very time-consuming since the PRDs solve the problem one at a time and update the system's operator (non-simultaneous). This approach could be time-consuming if many PRDs exist in the network. They proposed a faster semi-optimal algorithm to compensate for this issue, but the result slightly deviates from the equilibrium.

Mohsenian et al. [18] developed a games theoric approach between competitive resources. All the participants communicate with each other in the grid and are aware of each other's behavior (nonprivate). The goal is to reach Nash equilibrium between the users, which will be equivalent to reaching the global minimum when the objective is cost minimization. The issue with this method is that all users are connected to each other. This topology creates two issues:1) privacy: given that all users must know about everyone's behavior, each user's privacy is jeopardized. Also, it would open up more space for suspicious behavior, given that each user has access to more information. 2): The need for high bandwidth and more computation capacity. Sending each individual's data to all other users requires a costly infrastructure to send the data and a high bandwidth. At the same time, a delay on each connection link could reduce the speed of each iteration in the real world.

## 3. Proposed Method

### 3.1 Original Method (OM)

While the above approaches try to find the Nash equilibrium (optimal point) through a game theory framework, an alternative would be a comprehensive optimization problem containing the cost of electricity generation and the end-user PRD's local objective (that part of its cost of using electricity) as shown in Figure 1 (a). It must be noted that this is different from multi-objective optimization. The problem in Figure 1 (a) describes the joint optimization of the Unit commitment problem and PRDs self-benefiting problem. The goal is to find a price signal and energy bid (decision variables) that minimize the user payment and generation of electricity cost.

In a case where the operator is completely aware of the preference of each individual appliance. (Note: in this paper, we assume PRDs are rational agents, and their preference would be obtained from solving a personal optimization problem) Therefore by knowing how they solve an optimization problem with a given price, there would be an optimal answer where the generation costs and end-users local objectives were minimized simultaneously. In such an optimum situation, the grid operator and the end-users wanted to keep their generation/consumption (because they are at the optimal point or Nash- Equilibrium). However, such a problem in this form will be impossible to solve for two reasons: first, it will be a massive problem. Second, the grid operator needs to have all the information regarding the preference of each and every appliance at each house. In the next section, we propose a method to solve such a problem using Dantzig-Wolfe (DW) decomposition method.

### 3.2 Dantzig-Wolfe Representation of the OM

DW [32] was first proposed to solve large-scale linear programming problems. It breaks the large problem into one master problem and some subproblems. The subproblems are solved independently and simultaneously in a separate computer and report their preliminary results (at each iteration) to a computer with the master problem. At the same time, the Block-Angular structure of the OM highly spars. I.e., the problem of device $J$ is independent of device $J+1$. The problems are connected only through one linking constraint (the constraint that forces the model to equal the overall generation with overall consumption. The abovementioned structure of the OM makes it ideal for solving with DW.

Figure 1 (b) shows the summary of the main structure of this algorithm. It is an iterative algorithm until the desired accuracy is achieved or no further improvement happens in the results.

In the proposed method, the PRD's objectives are modeled as the subproblem of the DW, and the unit commitment objective act as the master problem of the DW. The solution to the master problem (unit commitment) would be the proposed load profile (energy bid) for each user and the corresponding price based on the marginal cost of electricity generation. The result of PRD's local optimization is load profiles based on the proposed price from the master problem (the energy profile they are willing to consume under the suggested price from unit commitment). This iterative procedure repeats until a point where the grid operator (which solved the unit commitment master problem) and the local PRDs (which solved subproblems) will not change their behavior based on the other side's proposal (energy bid from the PRDs side and price from unit commitment side) since the algorithm reached the optimal point.

Using DW to solve such a universal problem has numerous advantages 1) it is optimal. While it looks obvious, many appliance coordination algorithms proposed above can not find an optimal schedule for both sides of the market. One side needs to sacrifice for the other side. 2) it is fast. Since most of the optimization is done locally and simultaneously on the PRDs side, each iteration is rapid. Furthermore, the only information needed is the price signal and energy bid from unit commitment and PRDs sides, respectively. The simultaneous update is way faster than the algorithm proposed in[12], [18], that one appliance solves the problem at a time. 3) it is private. The appliances only share information regarding their consumption plan with the grid operator and only with the unit commitment side. This approach also prevents a distrustful coalition between the appliances from manipulating the master problem result based on their benefits. This method contrasts with the method proposed in [82], [84], where all appliances must share information. 4) it is a market-based algorithm. The utmost goal of this algorithm is to find the balance between supply and demand through a market. This paves the way for amore significant incorporation of distributed generations (including renewables) into the power system.

The chapter's structure is as follows: in capture 4 the fundamental principle of DW is explained. In capture 5.1, a simplified united commitment model for the grid operator and self-benefiting algorithms for appliances (EVs and EWHs) are proposed. Capture 6 discussed the results of implementing this algorithm.

## 4. Joint Unit Commitment and PRD cost minimization problem

### 4.1 Power System Operation with PRDs

Some key elements of the unit commitment problem for electric power systems with PRDs can be written as

$$\min_{\mathbf{g}_i,\mathbf{d}_j,\mathbf{z}_j,i\in G, j\in D} \sum_{i\in G} c_i(\mathbf{g}_i) + \text{(other costs)} - \sum_{j\in D} b_j(\mathbf{d}_j, \mathbf{z}_j)$$
$$\text{s.t.} \sum_{i\in G} \mathbf{g}_i + \mathbf{g}_0 = \sum_{j\in D} \mathbf{d}_j + \mathbf{d}_0 \quad (1)$$
(and other system-level constraints)
(and device-level constraints)

In this model, $G$ is the set of all controllable power plants, and $D$ is the set of price-responsive devices. The main decision variables are the power output from plant $i$ ($\mathbf{g}_i$), and the amount of power to be consumed by device $j$ ($\mathbf{d}_j$). These are vectors with one element for each hour of the next day. $\mathbf{z}_j$ is a vector containing all the other decisions that must be made by device j, e.g., temperature or charge levels for each hour. Together, $\mathbf{d}_j$ and $\mathbf{z}_j$ constitute a complete operating plan for device $j$ for the next day. The function ci shows the cost of producing power in plant $i$. The function $b_j$ shows the benefits to the owner of device $j$ when it follows the specified operating plan, converted to dollar form, i.e., the most the customer would be willing to pay to follow this plan. The only constraint shown is that the total generation scheduled for the next day, plus the noncontrollable generation $\mathbf{g}_0$ (e.g., distributed solar power), must equal the total demand from price-responsive devices plus the non-price-responsive market $\mathbf{d}_0$.

For brevity, we have omitted numerous additional terms: costs for starting, stopping, and running power plants; constraints on minimum and maximum output levels for each plant; minimum up- and down-times; and transmission line loading during regular operation or contingencies [34]–[40]. These could be included in future work.

If we split problem (1) into a power production cost function $C(\mathbf{D})$ and a consumption benefit function B(**D**) as follows:

$$C(\mathbf{D}) = \min_{\mathbf{g}_i, i\in GK} \sum_{i\in G} c_i(\mathbf{g}_i) + \text{(other system-level costs)}$$
$$\text{such that} \sum_{i\in G} \mathbf{g}_i + \mathbf{g}_0 = \mathbf{D} + \mathbf{d}_0 \quad (2)$$
(and other system-level constraints)

$$B(\mathbf{D}) = \max_{\mathbf{d}_j,\mathbf{z}_j, j\in D} \sum_{j\in D} b_j(\mathbf{d}_j, \mathbf{z}_j)$$
$$\text{such that} \sum_{j\in D} \mathbf{d}_j = \mathbf{D} \quad (3)$$
(device-level constraints)

then problem (1) becomes simply

$$\min_{\mathbf{D}} C(\mathbf{D}) - B(\mathbf{D}) \qquad (4)$$

In other words, the system operator must find a consumption plan **D** that can be served at the lowest net cost, where net cost is defined as the system's production cost function C(**D**) minus the customers' benefit B(**D**).

## 5. Dantzig-Wolfe-Based Coordination Mechanism

For large power systems, it may be challenging to convey the details of the PRDs' benefit problems to a central coordinator. Even if the central coordinator had those details, the entire unit commitment problem with millions of PRDs would likely be too large to solve on a single computer. In this paper, we instead present a distributed, iterative coordination mechanism based on the Dantzig-Wolfe decomposition of the unit commitment problem, which relies on PRDs to calculate their own parts of this problem.

The unit commitment problem with PRDs has two properties that make this possible. First, the PRDs' local optimization problems can be separated from the main unit commitment problem, except for the requirement that total production equals total consumption. Second, the PRDs' private optimization problems are generally convex (often linear), e.g., choosing hourly consumption levels to recharge a battery overnight fully. These properties are true for the individual PRDs' optimization problems and also for the overall consumption benefit problem, which is the sum of all the independent problems for each PRD.

Dantzig-Wolfe decomposition is a well-known iterative technique for solving optimization problems of this form that are too large to represent directly on a single computer [32], [33], [41]–[43]. As such, it is a good choice for the unit commitment problem with millions of PRDs. Dantzig-Wolfe decomposition also comes with attractive guarantees: it provides a better solution with each iteration; convergence occurs in finite time with linear subproblems; at each iteration the difference between primal and dual objective values provides a measure of the distance from optimality; and if interrupted before convergence, it will always provide a feasible solution.

Dantzig-Wolfe decomposition relies on two facts about convex programs: (1) any feasible solution can be expressed as a convex combination (weighted average) of the extreme points of the program (i.e., the feasible region is all the space inside the convex hull of a collection of points known as the "extreme points" of the problem), and (2) the program's objective function (net cost) can always be expressed as a convex combination of the objective value at the extreme points (i.e., the objective function is the convex hull of the objective values at those same points).

These facts allow us to reframe the consumption benefit function in terms of weights are applied to the extreme points of the consumption problem:

$$B(\mathbf{D}) = \max_{w_k} \sum_{k \in K} w_k b_k$$
$$\text{s.t.} \quad \mathbf{D} = \sum_{k \in K} w_k \mathbf{d}_k \qquad (5)$$
$$w_k \geq 0, \quad \forall k \in K$$
$$\sum_{k \in K} w_k = 1$$

where $K$ is the set of all feasible points, k is an index in $K$, $\mathbf{d}_k$ is the demand at point $k$, $b_k$ is the value of the benefit function at that point, i.e., $b_k = B(\mathbf{d}_k)$, and $w_k$ is the weight given to point $k$. With this in mind, we can rewrite problem (4) as the Dantzig-Wolfe master problem:

$$\min_{w_k: k \in K} C(\mathbf{D}) - B(\mathbf{D}) = C\left(\sum_{k \in K} w_k \mathbf{d}_k\right) - \sum_{k \in K} w_k b_k$$
$$\text{s.t.} \sum_{k \in K} w_k = 1 \qquad (6)$$

In this formulation, the master problem consists of constructing a consumption plan **D**—a convex combination of extreme points of the consumption problem—that can be served at the lowest net cost, where net cost is defined as the system's production cost function C(**D**) minus the customers' benefit B(**D**).

It is not usually possible to enumerate all the members of the set *K*. Instead, Dantzig-Wolfe decomposition builds up a subset *K'* that defines the feasible space near an optimal solution [32]. This subset is created by iterating between the master problem, as shown in (6), and the demand-side sub-problem:

$$\min_{\mathbf{D}} \mathbf{p}^T \mathbf{D} - B(\mathbf{D}) \qquad (7)$$

Here, **p** is a vector of hourly electricity prices. These are set equal to the marginal cost of serving additional load in the master problem (6) (e.g., dual values of the load-serving constraint) during the previous iteration. When the consumption subproblem (7) is solved with these prices, it finds the best possible consumption plan **D** at these prices. This solution must be at an extreme point of the consumption problem, so this provides another point $\mathbf{d}_k$ and benefit value $b_k = B(\mathbf{d}_k)$ to add to the set of extreme points $K'$.

This continues until there is no new extreme point that can improve the solution to the problem, at which point the issue has converged on the optimal consumption plan [32]. If $B(\mathbf{D})$ is a linear program (or equivalent to one), it has a finite number of extreme points so that convergence will be achieved in finite time. On the other hand, nonlinear, convex programs are equivalent to linear programs with infinitely many extreme points. In this case, the solution will improve at each step, asymptotically approaching the optimum.

Consequently, the unit commitment problem with PRDs can be solved as follows. In each iteration, the master problem (6) chooses the best combination of all the previously reported demand vectors and offers tentative prices to the consumption subproblem; then consumption subproblem offers a demand vector that can be served at equal or lower net cost (cost minus benefit). Specifically,

1. The central coordinator makes an initial estimate of prices $\mathbf{p}_{(0)}$.
2. The consumption subproblem (7) is solved on a distributed basis:
    a. The central coordinator passes the price vector $\mathbf{p}_k$ to the PRDs.
    b. All PRDs solve their individual subproblems and return $\mathbf{d}_{jk}$ and $b_{jk}$. These are their consumption plan and an estimate of the benefit of this plan (i.e., the amount they'd be willing to pay to follow this schedule). For simplicity or privacy protection, the benefit can be specified with an arbitrary offset that is constant over all iterations.[2]
    c. The $\mathbf{d}_{j,k}$ and $b_{j,k}$ vectors from individual PRDs are summed to create a system-wide consumption plan $\mathbf{d}_k$ and benefit value $b_k$.
3. The central coordinator adds the new demand profile $\mathbf{d}_k$ with benefit value $b_k$ to the set of extreme points $K'$ and solves the master problem (6), choosing a new combination of previously reported extreme points. If an optimality threshold is reached or an iteration limit has been reached, then the process terminates. Otherwise, the objective and pricing have been slightly improved, and the central coordinator sets new prices $\mathbf{p}_{k+1}$ equal to the dual values of the load-serving constraints. Then the process returns to step 2.
4. At the end of the iterations, the system coordinator reports the final nonzero weights $w_k$ to the PRDs (or possibly to the home or feeder coordinator). These are then multiplied by the load bids previously received from individual PRDs. These load levels are then assigned as the amount of power to be used by each PRD. If the problem converges, these load vectors are optimal for individual PRDs and the whole system at the final prices. If the process stops short of convergence, then the final price and load vectors are still known to be feasible and an improvement over the starting point for both the system and the individual PRDs.

Importantly, this mechanism can choose any mixture of previous demand bids by the PRDs, rather than being confined to the exact load profiles that have been bid. With highly price-sensitive devices, this has significant advantages over approaches that attempt to find just the right price to induce the right demand: (1) If very price-sensitive PRDs such as EVs or customer-sited batteries are on the margin in multiple hours, they will arbitrage away any price differences; this occurs because even infinitesimal price differences between hours would cause large swings of load between those hours. Consequently, the efficient prices may be flat over the period when these devices are on the margin, and will fail to send a signal to achieve a particular shape of load (e.g., valley-filling or renewable-following). Instead, a quantity signal is also needed, such as the one provided here. (2) In practice, price-based coordination mechanisms often work by adjusting prices in small steps until supply and demand are balanced. However, that is not possible with infinitely price-sensitive devices, since even infinitesimal variations will cause all the load to swing in one direction or the other; this causes those methods to oscillate without converging on a single right price vector (and as noted above, even if they found the right price vector, that alone would not induce the right consumption vector). In our work, described below, the Dantzig-Wolfe mechanism performs well for highly-price-sensitive devices, choosing a mix of demand vectors from either side of the equilibrium point and quickly moving toward convergence.

---

2 For devices where the direct benefit does not vary among all feasible operating plans (e.g., an electric vehicle that always charges fully regardless of prices), the benefit of any feasible plan could be reported as zero. For devices where the benefits vary from one plan to another, the value of the benefit function should change by the same amount.

It should also be noted that the consumption subproblem can naturally be divided into separate subproblems for each device. Then the main subproblem can be solved by passing the prices to all the individual devices, letting them solve their portion of the subproblem, and summing the load vectors and benefit values that they return. This could allow for highly parallelized computation and efficient communication in a real-world implementation. For example, the master problem could be solved by a central coordinator, while nodal coordinators manage each distribution circuit, and home coordinators manage the individual PRDs within each home. Devices at each level of the hierarchy simply announce price vectors to the devices or coordinators below them and then sum the demand vectors and benefits that they receive back. This allows data to be condensed by several orders of magnitude at each level, so very little bandwidth is needed at each level (just enough to communicate with a few dozen elements at the next lower level). It is also worth noting that this technique can incorporate highly heterogenous PRDs without any modification, i.e., any device that has a convex (rational) response to prices.

## 5.1 Price Responsive Device (PRD) Models

In the literature, loads are categorized as non-shiftable, shiftable and thermostatically controlled appliances (TCA) [44]. non-shiftable (uninterruptible) are loads that are the one that energy must be delivered to them instantly. TVs or stoves are from the non-shiftable group. The interruptible are appliances that instant energy usage is not needed since either they have batteries or are needed once per day or week while urgent usage is not needed. EVs, dishwashers and driers are from this category. The TCAs are the ones where the energy consumption is a function of temperature (ambient temperature or user's desired temperature). The working temperature (desired) could be affected by the ambient temperature, or it could be affected by input/output of the energy through consumption or added energy from the heating/cooling elements respectively. In many works such as [18] only non-interruptible and interruptible loads are considered for the task of energy management. However, since non-shiftable loads have no response to any control command (either from direct load control or optimization), they could be removed from the problem without losing generality. Since it is not possible to show the dynamic response of all types of loads in one paper and the fact that their behavior is similar to some extent, EVs from shiftable loads and EWHs from TCAs are selected to represent a sample of load profile. This section covers the self-benefiting model of PRDs that participate in the market bidding. i.e., the EWH and EV.

### 5.1.1 Electric Water Heater Model

The following shows a self-benefiting model of EWH that aims to minimize its cost. It is a day-ahead optimization implemented as a linear program based on the model developed in Section 1 and [5]. It is dual-objective, i.e., the model tries to find a balance between cost saving and discomfort. For this study, we assume perfect foresight of hot water requirements and optimize power consumption accordingly. In future work, we could consider an EWH that buys vectors of power as shown here, based on a stochastic forecast of hot water usage, as discussed in Section 1 and [5].

Equation (8) shows the dual-objective cost function which minimizes the cost from use of electricity and the discomfort cost (underheated water) that could happen from using this program.

$$\min_{E^{\text{in}} \in \mathbb{R}^{24}} \sum_{h \in I} p_h^{\text{p}} E_h^{\text{in}} + p^s E_h^{\text{short}} \quad (8)$$

Equation (9) models the thermodynamic behavior of EWH.

$$T_{d,h}^{\text{tank}} = T_{h-1}^{\text{tank}} + \frac{(E_h^{\text{in}} - E_h^{\text{w}})}{c^{\text{tank}}} - r^{\text{loss}} \cdot (T_h^{\text{tank}} - t_h^{\text{amb}}) \quad (9)$$

In (10), the maximum energy that can be drawn from the grid is limited to the EWH energy input capacity.

$$E_h^{\text{in}} \leq e^{\max} \quad (10)$$

Equation (11) and (12) model the temperature and energy shortfall respectively. They calculate the underheated water that would be caused by the plan.

$$T_h^{\text{short}} \geq t^{\min} - T_h^{\text{tank}} \quad (11)$$

$$E_h^{\text{short}} = e_h^{\text{des}} \frac{T_h^{\text{short}}}{(t^{\min} - t_h^{\text{in}})} \quad (12)$$

After solving this problem, PRD number j reports its consumption vector during iteration k as $\mathbf{d}_{j,k} = [E_0^{\text{in}}, E_1^{\text{in}}, E_2^{\text{in}}, \ldots, E_{23}^{\text{in}}]$. It also calculates the benefit $b_{j,k}$ as being equal to the shortfall penalty times the quantity of hot water consumed (this may vary from one iteration to the next, as the plan adapts to different electricity prices). By definition, this is the amount that the customer would be willing to pay for the hot water they will receive, and by extension, for the power that will be used to produce this hot water.

### 5.1.2 Electric Vehicle Model

The electric vehicles follow a simple cost-minimizing algorithm. The user defines the hours that the EV can be charged (parking hours) in advance. These hours are expressed through the maximum charging rate per hour. i.e., $e_h^{max}$ for hour $h$. The $e^{des}$ is the energy needed by the end of the charging window (i.e., the total energy required for the day). Then, the role of the following cost-minimizing programming is to find the cheapest hours to pick to charge the vehicle.

The cost function is shown in equation (13). The objective is simply to minimize the user's electricity cost.

$$\text{the } \underset{E^{\text{in}} \in \mathbb{R}^{24}}{\operatorname{anmin}} \sum_{h \in I} p_h^{\text{p}} E_h^{\text{in}} \quad (13)$$

In this equation, $I$ is the set of all hours in the coming day and $p_h^{\text{p}}$ is the (tentative) price for power during each of those hours, and $E_h^{in}$ is the decision variable showing the amount of energy to use for charging during each hour.

The charging window is modeled through the equation (14):

$$E_h^{in} \leq e_h^{max} \quad (14)$$

The total energy added to the battery through the charging window must be equal to the total energy needed. Equation (15) shows this constraint.

$$\sum_{h \in I} E_h^{\text{in}} = e^{\text{des}} \quad (15)$$

We use a simple algorithm to solve this linear program:

- Select hours of the day when charging is possible ($e_h^{max} > 0$).
- Sort these hours from lowest to the highest price, then from early to late as a tie-breaker.
- During each hour, starting with the first in the list, add enough charge to reach $e^{\text{des}}$, or charge at rate $e_h^{max}$, whichever is less.

After choosing the optimal charging plan, the EV model reports back the consumption vector **the $\mathbf{d}_{j,k} = [E_0^{\text{in}}, E_1^{\text{in}}, E_2^{\text{in}}, \ldots, E_{23}^{\text{in}}]$**. When using this algorithm, the EV always obtains a full charge, so the benefit to the EV is the same regardless of the charging plan. So the benefit $b_{j,k}$ is reported as zero for each EV during each iteration.

## 5.2 Aggregated Consumption Benefit Problem

For the Dantzig-Wolfe iteration described in this paper, the consumption subproblem (7) is regarded as a single problem encompassing all PRDs in the system. That is achieved simply by offering the same price vector $\mathbf{p}^p = \mathbf{p}_k$ to all the water heater and electric vehicle models, solving the model for each individual PRD, then summing the consumption plans for all the PRDs:

$$\mathbf{d}_k = \sum_{j \in J} \mathbf{d}_{j,k}, \text{ and } b_k = \sum_{j \in J} b_{j,k} \quad (16)$$

These are then reported to the central coordinator.

After the last iteration of the master problem, the final weights $w_k$ from the master problem (6) are assigned back to the PRDs, based on each device's prior bids:

$$\mathbf{d}_j^* = \sum_{k=1}^{N} w_k \mathbf{d}_{j,k} \quad (17)$$

This could be done by reporting the weights to the devices themselves or by having the home or circuit coordinator calculate $\mathbf{d}_j^*$ and report it to device $j$ as the amount of power it has purchased.

With this assignment, the total power consumed by all PRDs matches the total expected/constructed in the system-level master problem:

$$\mathbf{D} = \sum_{k=1}^{N} w_k \mathbf{d}_k = \sum_{k=1}^{N} w_k \sum_{j \in D} \mathbf{d}_{j,k}$$
$$= \sum_{j \in D} \sum_{k=1}^{N} w_k \mathbf{d}_{j(k)} = \sum_{j \in D} \mathbf{d}_j^* \quad (18)$$

This framework assumes that the PRDs all have convex responses to price. It foregoes the option of assigning different weights to the extreme points (bids) elicited from the individual devices. In future work, we could incorporate devices with integer decisions (e.g., dishwashers or clothes washers that must run an entire cycle once they start). One promising option for this would be to fine-tune the weights for individual PRDs at the feeder (neighborhood) level, requiring integer weights for the integer PRDs, and adjusting the weights for all the PRDs on the feeder so that the total for the feeder matches the total load requested by the system as closely as possible.

### 5.3 Power System Model

For this paper, we use a simple electricity production model. We ignore non-dispatchable generation $\mathbf{g}_0$ and non-price-responsive loads $\mathbf{d}_0$, so the load-balancing constraint of equation (2) becomes

$$\sum_{i \in G} \mathbf{g}_i = \mathbf{D} = \sum_{k \in K} w_k \mathbf{d}_k \quad (19)$$

We also assume that the system has a quadratic production cost so that the minimum cost of producing an amount of power $\mathbf{D}$ from the available generators is

$$C(\mathbf{D}) = \sum_{h=1}^{24} a D_h^2 \quad (20)$$

where $D_h$ is the element of the demand vector $\mathbf{D}$ for hour $h$.

These assumptions create a simple system with easily calculated marginal costs ($2aD_h$) each hour. However, the approach shown in this paper is general enough to apply to any convex supply side, e.g., a linear program where marginal costs are the dual values of the hourly load-balancing constraints.

### 5.4 Optimality Gap

By step $N$ of the Dantzig-Wolfe iteration, the PRDs in the system have revealed the following extreme points (feasible consumption quantity vectors):

$$\mathbf{d}_k, \quad k \in K' = 1..N \quad (21)$$

Where $N$ is the number of iterations/bids we've done, and $\mathbf{d}_k$ is the total consumption bid from all PRDs during step $k$. The benefit to the customers for each extreme point (the maximum they would be willing to pay for that vector, possibly shifted by a constant offset) is

$$b_k, \quad k \in K' = 1..N \quad (22)$$

where $b_k$ is a scalar, the sum of all the benefits reported by the individual PRDs.

Then, any weighted combination of these bids is possible because the feasible regions for the EV and EWH models are convex. So, the master problem constructs a consumption plan (load vector) that is a weighted sum of the previous bids:

$$\mathbf{D} = \sum_{k=1}^{N} w_k \mathbf{d}_k \quad (23)$$

where $\mathbf{D}$ is the constructed bid (vector, one element for each hour $h$).

The total benefit of this constructed bid is $B(\mathbf{D})$, where

$$B(\mathbf{D}) \geq B'(\mathbf{D}) = \sum_{i=1}^{N} w_k b_k \quad (24)$$

The central coordinator doesn't know $B(\mathbf{D})$ directly, but over multiple iterations, the reduced-form representation $B'(\mathbf{D})$ will converge to the actual value of $B(\mathbf{D})$ [32]. As noted previously, this will occur in finite time if the PRD subproblems are linear programs (as they are for this paper), and will occur asymptotically if the PRD problems are convex but nonlinear [108].

Dantzig-Wolfe decomposition gives an optimality gap at each iteration, which can be seen as the difference between two measures: $S^{\text{best}}$ (the best possible value of the objective function) and $S^{\text{known}}$ (the best objective value found so far).

During round k, the master problem minimizes $C(\mathbf{D}) - B'(\mathbf{D})$. The value of the objective function at this point is

$$S^{\text{known}} = C(\mathbf{D}_k) - B'(\mathbf{D}_k) \quad (25)$$

where $\mathbf{D}_k$ indicates the value of $\mathbf{D}$ selected by the master problem in round k.

$S^{\text{known}}$ is a lower bound on the objective function because it is known to be achievable. $B'(\mathbf{D}_k)$ in the

master problem is conservative. It is a weighted combination of points on the true $B(\mathbf{D})$ function, so it must be equal to or below the true $B(\mathbf{D}_k)$ with loads $\mathbf{D}_k$, since $B(\mathbf{D}_k)$ is convex. $C(\mathbf{D}_k)$ is also achievable because it is the actual production cost function.

After solving the master problem in round $k$, the central coordinator offers tentative prices $\mathbf{p}_k$ to the customers, and they propose to buy a total amount of power each hour $\mathbf{d}_k$. They also report that this power gives them the benefit of $b_k$.

The subproblem solution $\mathbf{d}_k$ (with benefit $b_k$) is optimal, which means that all other values of $\mathbf{d}$ would give equal or worse objective values, i.e.,

$$B(\mathbf{d}) - \mathbf{p}_k \cdot \mathbf{d} \leq b_k - \mathbf{p}_k \cdot \mathbf{d}_k \qquad (26)$$

for all $\mathbf{d}$, so

$$B(\mathbf{d}) \leq b_k - \mathbf{p}_k \cdot \mathbf{d}_k + \mathbf{p}_k \cdot \mathbf{d} \qquad (27)$$

The master problem has solution $\mathbf{D}_k$, with production cost $C(\mathbf{D}_k)$ and incremental production cost $\mathbf{p}_k$, i.e., $\mathbf{p}_k$ is the gradient of $C(\mathbf{d})$ at $\mathbf{D}_k$. Since $C(\mathbf{d})$ is convex downward, it must lie above the plane that is tangent at this point, so

$$C(\mathbf{d}) \geq C(\mathbf{D}_k) + \mathbf{p}_k \cdot (\mathbf{d} - \mathbf{D}_k) \qquad (28)$$

for all possible demand vectors d.

Subtracting equation (27) from (28) gives

$$C(\mathbf{D}) - B(\mathbf{D}) \geq C(\mathbf{D}_k) - \mathbf{p}_k \cdot \mathbf{D}_k - b_k + \mathbf{p}_k \cdot \mathbf{d}_k \quad (29)$$

so

$$S^{\text{best}} = C(\mathbf{D}_k) - \mathbf{p}_k \cdot \mathbf{D}_k - b_k + \mathbf{p}_k \cdot \mathbf{d}_k \qquad (30)$$

Then the optimality gap is

$$\begin{aligned} S^{\text{known}} - S^{\text{best}} &= C(\mathbf{D}_k) - B'(\mathbf{D}_k) \\ &\quad - C(\mathbf{D}_k) + \mathbf{p}_k \cdot \mathbf{D}_k + b_k - \mathbf{p}_k \cdot \mathbf{d}_k \\ &= [b_k - \mathbf{p}_k \cdot \mathbf{d}_k] - [B'(\mathbf{D}_k) - \mathbf{p}_k \cdot \mathbf{D}_k] \end{aligned} \qquad (31)$$

For this paper, we always performed 24 iterations and only used the optimality gap as an informative diagnostic. However, in future work, we could iterate until the gap is below a fixed limit or a fixed fraction of the direct costs $C(\mathbf{D}_k)$.

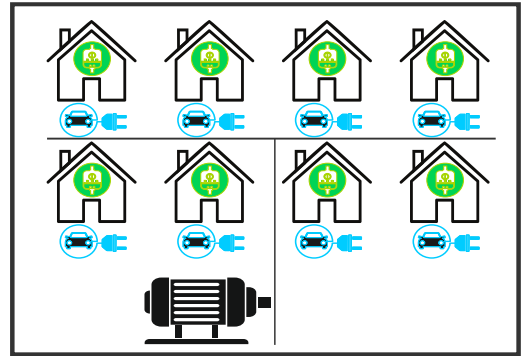

*Figure 2 Grid structure*

## 6. Results

### 6.1 Network structure and problem assumption

The structure of the smart grid is shown in Figure 2. This network serves eight households (a total of 16 appliances). All the loads are connected to the grid operator through a common line. Without losing the generality, it is assumed that all units are being fed through one distributed generator (DG) unit that feeds all the loads. Since the only energy source is one dispatchable unit and no renewable energy that can affect generation cost, the marginal cost for all hours is the same. Also, power quality constraints such as voltage and frequency regulations are not considered in this unit commitment problem, and the focus is on economic dispatch.

### 6.2 Load type:

In this work, it is assumed that the generator belongs to all consumers. This means the operator acts only as a coordinator and does not seek its benefit. Each customer pays the operational cost of using the generator unit, which is equal to their consumption profile times the price proposed by the coordinator. Each household ONLY sends its energy bid to the coordinator, and no neighbor is aware of other individuals' consumption patterns.

Each household responds to the day-ahead price signal and sends the aggregate energy consumption of all devices in its network (EV and EWH). As mentioned above, the loads that can participate in the bidding process are both shiftable and TCA loads. To narrow the scope, EVs are selected among the shiftable loads, and electric water heaters are selected from the TCA loads. Therefore, in this work, each household responds to price signals to optimize the performance of its electric vehicle and electric water heater. All other loads (including other non-shiftable, shiftable, and TCAs) are removed from the load profile to simplify the problem.

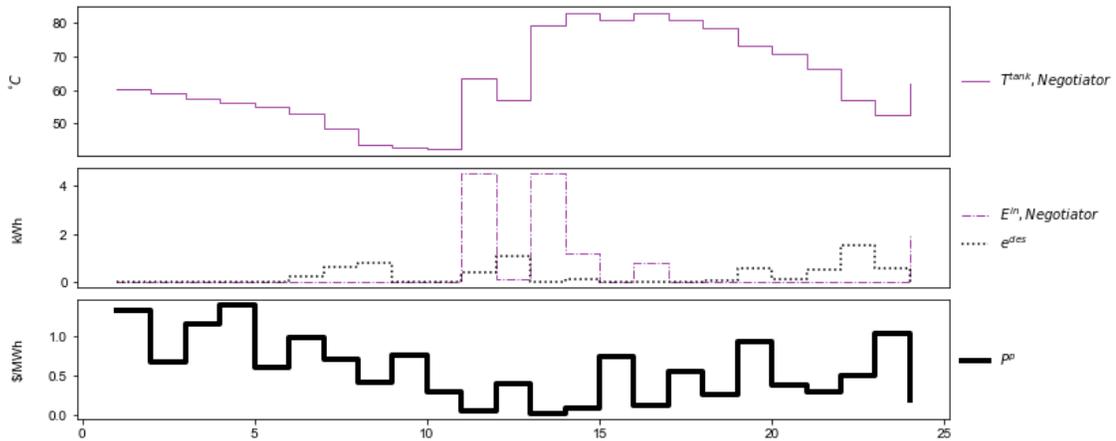

*Figure 3 EWH response to the broadcasted price*

In this hypothetical scenario where the generation cost is equal for all hours (since no renewable energy generation exists) and only demand-responsive loads exist, the ideal aggregate load and price signal would be flat for all intervals. However, the current model could easily be changed to include non-interruptible loads of renewable energy sources.

## 6.3 Individual appliances respond to the price signal and their benefit

As mentioned in the previous section, only two types of appliances are considered for this simulation. Also, it was mentioned that loads are categorized into three main types: 1-non-shiftable, 2-shiftable, and 3-TCAs. Non-shiftable loads don't respond to price, and their loads are fixed. Therefore, to simplify and focus to the claims of this paper, non-shiftable loads are removed from the load profile. In this section, we focus on the performance of an individual EV and EWH and show how they selfishly try to maximize their benefit by responding to the price signal by running the optimizations that are proposed in section 5.1.

### 6.3.1 EWH Response

Figure 3 shows the EWH response to the price signal sent from the coordinator for one round of optimization. Based on this price and prediction of user energy consumption, the EWH finds the optimal answer (self-benefiting). The energy usage that is the decision variable of EWH is sent to the aggregator as the energy bid on behalf of the EWH. The price signal is shown in black in the third section of the figure. At the same time, the EWH has a prediction for the energy withdrawal for the user. This prediction can come from analyzing historical data and using the machine learning method to obtain this prediction. The prediction in this work is considered perfect (not probabilistic) and is shown in the middle section of the figure with dotted black lines. The main variables that resulted from solving the optimization problem that was explained in section 5.1.1. Section one shows the actual temperature of the water heater based on the predictions. The pink dotted line in the middle section represents the energy input that EWH plans to use. This is the variable that is sent to the coordinator after each round of solving the problem. A more detailed discussion of the EWH optimization model is provided in [5].

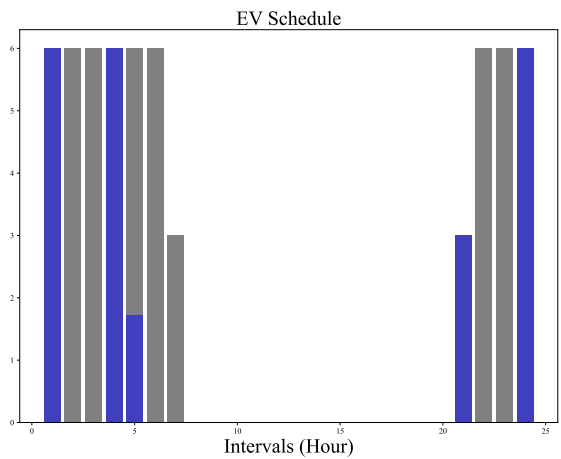

*Figure 4 EV response to broadcasted price*

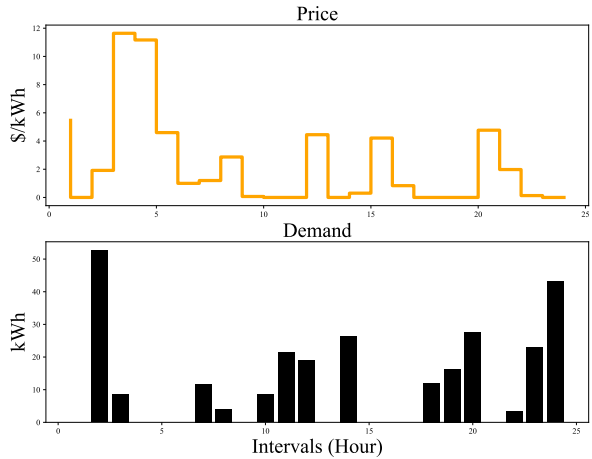

*Figure 5 initial load & price profile*

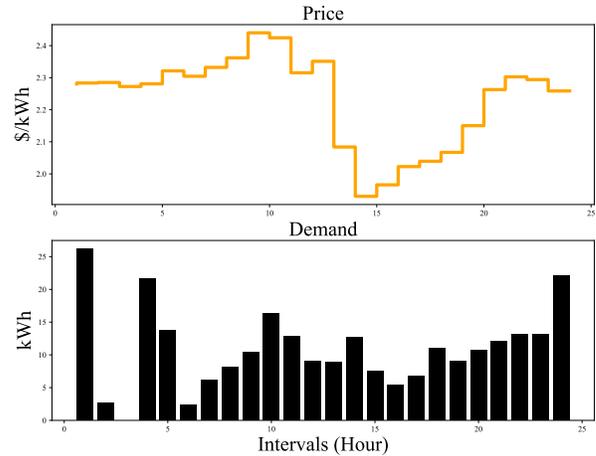

*Figure 6 final load & price profile*

### 6.3.2 EV Response

Figure 4 shows the response of the electric vehicle to the price signal. The gray bars show the energy the EV can absorb at each hour. The times that availability is zero means that the vehicle is operational and can't be charged at those intervals. The blue bars show how much energy EV will absorb based on the optimization result explained in section 6.3.2. Similar to the EWH response, the energy input, which is the decision variable of EVs (blue bars), is reported to the coordinator as the energy bid of the EV. For example, this particular EV absorbed energy at full capacity in hours 1, 4, 21, and 24 and worked with one-third of its capacity at hour 5.

### 6.4 Price and Aggregate Demand Profile

in a problem where marginal cost is the same for all hours (since there is no renewable energy source for this problem) and there is no fixed load, the ultimate ideal scenario for aggregate load and price profile is a flat line. i.e., the Peak to Average Ratio (PAR) is one. Therefore, to evaluate the performance of the DW algorithm, we must see how far from the flat curve the resulting aggregate load and price profile are from the flat lines.

Figure 5 and Figure 6 show the price signal and aggregated load (in response to that price) at step 0 and the last step, respectively. i.e., each appliance solves its local self-benefiting optimizations in response to this price and reports back its scheduled energy usage as an energy bid. As can be observed, the initial demand is accumulated only during specific hours (as mentioned before, only EVs and EWHs are used in this work). So, it means that the price generated by the expected marginal cost of electricity generation (based on energy consumption) will no longer represent the marginal cost of generation. The reason is that the PRDs change their behavior to maximize their benefit (which means they will shift their consumption to cheap hours). This accumulation causes a high generation cost for the hours that were initially expected to be sparsely used. As it can be seen, the initial price is high between 3-6 AM. This sparks a low energy consumption in these expensive hours. While a 54 KWh is recorded as a peak hour that causes a high. So, this response means that if the coordinator price generation is only based on the current expected energy consumption, the algorithm never converges, and a flip-flop response is observed.

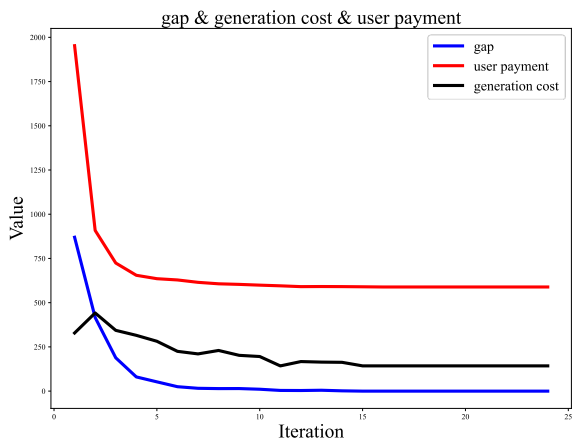

*Figure 7 Gap & Generation cost & user Payment*

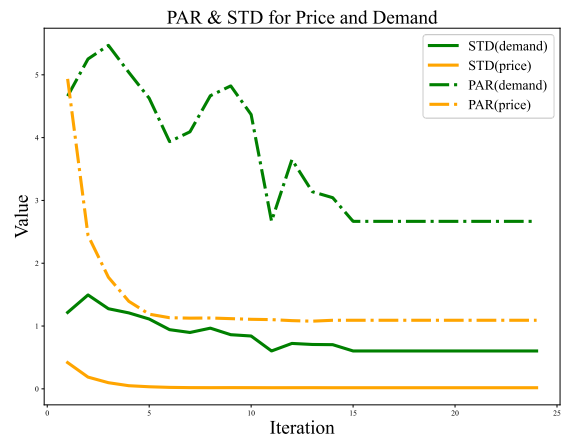

*Figure 8 PAR & STD for price and aggregate demand*

However, this point is where the importance DW algorithm shows up. The DW finds the optimal point between the previously reported bits at each optimization step. For example, in the third step of the optimization, by using the interpolation between the first bid and the second bit. This guarantees that the result of the third step is more optimal than the first two steps. This process is valid for other steps as well. For example, in the nth step, the algorithm has access to the model's behavior for the n-1 points. This means it can find an answer for the nth step that is at least as good as the previous step or better. This algorithm is repeated until no further improvement happens in the system.

Figure 6 shows the price and aggregate demand when the iterative algorithm finishes the optimization. The PAR reduces significantly, and the peak load is reduced from 54 KW to 26 KW. It must be noted that all loads are price responsive in this simulation. In a more realistic scenario, where part of the loads is non-responsive to the price, the amount of PAR will be less than the time when all loads are price responsive. As shown in Figure 6, the PAR for the price signal reduces significantly from 4.7 to 2.6.

### 6.5 Progress through iteration

As mentioned before, in each iteration, the coordinator sends the price signal to each appliance in each house. Then each device maximizes its benefit by solving an optimization problem locally given the broadcasted price signal. When the optimization reaches a certain level, when the price signal is broadcasted, none of the appliances change their bid. This shows that the network reached the global optimum or the Nash equilibrium. Figure 5 shows the progress of the DW through iterations. In the initial iteration of iterations, the variation is high for both the price and aggregated demand. The PAR for the price and aggregate demand is around 4.7 at this point. As the iterations progress, the PAR and standard deviation (STD) reduce. At iteration 5 and after, the PAR reduces to approximately 1.1. however, this variation is significant enough to induce a response to the loads to change their consumption. At iteration 15, no changes happen to the reaction of loads to the price. i.e., the algorithm reaches a point where no appliance wants to change its consumption plan (definition of Nash Equilibrium). This shows that the algorithm reached its primary goal: propose a price/energy bid to each individual user where they don't want to change their behavior. The optimality of the algorithm is discussed in the next section.

### 6.6 User Payment & Generation Cost & Optimality Gap

The generation cost and aggregate user payments are shown in Figure 7 in red and black, respectively. User payment is the sum of electricity prices times the aggregate consumption for that interval. The generation cost is calculated using the formula (20). As shown in Figure 7, generation cost and user payment are both high at the start of the algorithm. As the negotiation (iterations of the algorithm) goes forward, the cost of generation and user payment decreases. When the price signal is announced, none of the appliances change their consumption plans (they reach the Nash Equilibrium if we talk in terms of game theory). The gap, defined through formula (28), is shown in blue in Figure 7. The gap shows the distance between the best possible answer and the best-known answer. As the algorithm progress, the algorithm updates the value of these two variables. At the later part of the optimization (iteration 15 for this specific simulation), the distance between these two variables (Gap) stays fixed, and no further improvement happens (optimal point). Alternatively, as was mentioned in the previous section, the algorithm reaches the Nash equilibrium at iteration 15.

## 7. Conclusion

With the increased share of renewables and distributed generation in power systems, there is more need for a coordination algorithm to find an equilibrium between generation and consumption. One effective way to promote the goals in the smartgrid paradigm is creating a market for these services. Therefore, in this paper, a market-based coordination algorithm was proposed. However, coordination of PRDs could be very challenging due to the rapid change of PRDs energy consumption to even infinitesimal price signal changes and the lack of information about each preference of PRDs. Even if there is such an attempt to gain this information, such a method will jeopardize users' privacy.

To solve this issue, first, a universal optimization problem that jointly minimizes the generation cost and PRDs payments. Solving such a problem guarantees reaching an optimal point (or Nash equilibrium). However, solving such a massive problem is impossible due to a lack of knowledge of the end user's behavior. Furthermore, even if such knowledge (being aware of each user's choice) exists, solving such a massive problem will be computationally expensive (if possible). Therefore, the is a need for an algorithm that can solve such a huge problem while it does not need the private information of the users to reach the optimal point.

First, by breaking the universal problem into master (solved by the operator) and subproblems (solved by appliances), there will not be any need to know the preferences of each appliance. Instead, the appliances will only announce their energy usage as the energy bid to the operator. Second, due to the distributed behavior of the optimization, the algorithm will run very fast at iteration. Third, the way the algorithm is modeled guarantees convergence in limited iterations. Solving subproblems simultaneously and in a distributed fashion makes it possible to coordinate thousands of devices without losing speed.

Considering all the above advantages for a distributed price-based coordination algorithm, an algorithm based on the Dantzig-Wolfe decomposition was proposed. As it was shown in the results section, this proposed method could reach the optimal point in just 15 steps while it minimized the cost for the distributed generation unit and the PRDs so that no part wanted to change their behavior (reaching a balance).

In this work, a simple unit commitment with one generator was proposed. The next step for this project would be a more detailed unit commitment model with generation and transmission constraints in the optimization model. Given the proposed algorithm's efficiency and speed, it will be possible to solve a significant unit commitment problem with PRDs efficiently and optimally one day.

## 8. Acknowledgment

This research was partly funded by grants from the Ulupono Initiative and Blue Planet Foundation.